\PassOptionsToPackage{table,svgnames}{xcolor}
\documentclass[conference]{IEEEtran}
\IEEEoverridecommandlockouts
\usepackage[utf8]{inputenc} 
\usepackage[T1]{fontenc}
\usepackage{newtxtext}
\usepackage{multicol}
\usepackage{colortbl} 
\usepackage[hyphens,spaces,obeyspaces]{url}
\usepackage{cite}
\usepackage{amsmath,amssymb,amsfonts}
\usepackage{subcaption}
\usepackage{algpseudocode}
\usepackage{bmpsize}
\usepackage{graphicx}
\usepackage{tikz}
\graphicspath{ {./images/} }
\usepackage{textcomp}
\usepackage{xcolor}
\usepackage{makecell}
\usepackage[linesnumbered,ruled,norelsize]{algorithm2e}
\makeatletter
\newcommand{\removelatexerror}{\let\@latex@error\@gobble}
\makeatother

\usepackage{mwe}

\SetCommentSty{mycommfont}

\def\BibTeX{{\rm B\kern-.05em{\sc i\kern-.025em b}\kern-.08em
   T\kern-.1667em\lower.7ex\hbox{E}\kern-.125emX}}

\begin{document}

\title{Many Field Packet Classification with Decomposition and Reinforcement Learning}

\author{\IEEEauthorblockN{Hasibul Jamil , Ning Yang, and Ning Weng,~\IEEEmembership{Senior Member,~IEEE}} 
\IEEEcompsocitemizethanks{This work was performed while H. Jamil was at Southern Illinois University Carbondale. Hasibul Jamil is currently affiliated with Data Intensive Distributed Computing (DiDC) Laboratory, CSE, University at Buffalo, NY, USA. (E-mail: mdhasibu@buffalo.edu)
Ning Yang is affiliated with the School of Computing, Southern Illinois University, Carbondale, IL, USA. (E-mail: nyang@siu.edu)
Ning Weng is affiliated with the School of Electrical, Computer, and Biomedical Engineering, Southern Illinois University, Carbondale, IL, USA. (E-mail: nweng@siu.edu)

}
}

\maketitle

\begin{abstract}
 Scalable packet classification is a key requirement to support scalable network applications like firewalls, intrusion detection, and differentiated services. With ever increasing in the line-rate in core networks, it becomes a great challenge to design a scalable packet classification solution using hand-tuned heuristics approaches. In this paper, we present a scalable learning-based packet classification engine by building an efficient data structure for different ruleset with many fields. Our method consists of the decomposition of fields into subsets and building separate decision trees on those subsets using a deep reinforcement learning procedure. To decompose given fields of a ruleset, we consider different grouping metrics like standard deviation of individual fields and introduce a novel metric called diversity index (DI). We examine different decomposition schemes and construct decision trees for each scheme using deep reinforcement learning and compare the results. The results show that the SD decomposition metrics results in 11.5\% faster than DI metrics, 25\% faster than random 2 and 40\% faster than random 1.  Furthermore, our learning-based selection method can be applied to varying rulesets due to its ruleset independence.
\end{abstract}


\section{Introduction}

 In Software defined networking (SDN) architecture, control plane and data plane are separated, and that allows control plane and data plane technologies to evolve individually without one restricting the other. The traffic steering function of SDN's data plane is programmable by installing a set of $(match, action)$ rules on switches using a centralized controller. Each incoming packet is matched to the ruleset and receives the corresponding actions. To support fine-granularity in actions, the packet classification in switches needs to check far more header fields compared to 5-tuple packet classification techniques in traditional network switches. These additional header fields increase the ruleset complexity and pose great challenges to designing scalable many-field packet classification solutions for high-performance OpenFlow switches.
Major existing algorithmic solutions to packet classification are decomposition-based techniques~\cite{Lakshman98high-speedpolicy-based}\cite{Baboescu01_ABV} and decision-tree-based techniques~\cite{gupta:99hicuts}. However most of these solutions are built on heuristics (e.g., increasing split entropy~\cite{gupta:99hicuts}, balancing splits with custom space measures~\cite{gupta:99hicuts}, special handling for wildcard rules~\cite{singh2003packet}) that fails to generalize the process of building a decision tree for a different set of rules. On the other hand, if these solutions are specifically tuned to exploit certain characteristics present in a given ruleset, those characteristics may not be present in another ruleset. As a result, this environment (i.e., ruleset) specific heuristics typically suffers from a sub-optimal performance. Another drawback of this hand-tuned heuristics is the absence of a global objective (e.g., tree depth or the number of nodes in the tree). Their decision making is often based on local information (the difference between the number of rules in the current node~\cite{Hsieh-manyfield-tnsm-19}, the number of different ranges in different dimensions~\cite{singh2003packet}). This local information is loosely related to the global objectives and that leads to their performance to be sub-optimal. To address the above-mentioned limits of heuristics-based solutions, a learning-based approach is required~\cite{Liang-NPC-2019}. The challenge will be how to design a learning-based method, which will result in a memory and performance scalable packet classification engine~\cite{jamil2020}.

In this paper, we demonstrate a hybrid decomposition and decision-tree construction mechanism with deep reinforcement learning (DRL) for a next-generation packet classification engine. We revisit both the decision tree and decomposition techniques for traditional 5-tuple packet classification and propose a hybrid algorithmic process for the state-of-the-art packet classification engine. We employ the decomposition-based idea, given a 12 field ruleset, we partition the ruleset into multiple subsets from each of which an optimized decision tree is built. We use a learning-based approach to generate a high-performance decision tree with a low memory footprint for any ruleset without relying on heuristics.  We only consider 12 fields in this study but we believe due to the modular nature of our proposed scheme, it is not limited to just 12 fields but could be employed for ruleset with a higher number of fields. Ruleset with more fields is deferred for future work. In summary, there are three main contributions of this work: 
\begin{itemize}
\item A many-field packet classification algorithm is proposed by adopting a hybrid of 	decomposition and decision tree schemes. 

\item Two decomposition metrics are proposed and evaluated to decompose ten fields 	to five fields to considering performance and memory storage.

\item A ruleset independent method is proposed to build decision tree by using deep 	reinforcement learning.

\end{itemize}

The rest of this paper is organized as follows. Section~\ref{sec:back_motive} states the problem we intend to solve and summarizes the background of the decision-tree as a representative packet classification algorithm. A briefly review of traditional 5-tuple and new many-field packet classification solutions are presented in Section~\ref{sec:relatedWork}. Section~\ref{sec:sysOverview} overviews the whole design methods. Section~\ref{sec:learningMethod} describes a learning-based approach to build an optimized decision tree. Section~\ref{sec:decomposition} describes different approaches to partition the ruleset into subsets to build an optimized band of decision trees. Results and their discussion are presented in Section~\ref{sec:results}. Section~\ref{sec:conclusion} concludes this paper.

\section{Background}\label{sec:back_motive}
\subsection{Problem Statement}
\begin{table*}
\caption[]{Example: OpenFlow 12-field packet classification rule set generated with classbench-ng \cite{classbench-ng}. $E$ as exact match field, $P$ as prefix match field, and $R$ as range match field, $\star$ denotes don't care values. Action ${act_i}$ could be forward a packet to a set of ports, drop the packet; Add, modify, or remove vlan ID or priority on a per-destination-port basis, modify the destination MAC address, Send the packet to the  controller (Packet In), receive the packet from the controller and send it to ports (packet out)}

\label{tab:openflowRuleset}
\centering
\tiny
\begin{tabular}{|c|c|c|c|c|c|c|c|c|c|c|c|c|c|}
\hline
Rule                      & \begin{tabular}[c]{@{}l@{}}network   \\ source\end{tabular}           & \begin{tabular}[c]{@{}l@{}}network   \\ destination\end{tabular}            & \begin{tabular}[c]{@{}l@{}}transport   \\ source\end{tabular}                    & \begin{tabular}[c]{@{}l@{}}transport   \\ destination\end{tabular} & \begin{tabular}[c]{@{}l@{}}IP   \\ protocol\end{tabular} & \begin{tabular}[c]{@{}l@{}}layer 2   \\ source\end{tabular}           & \begin{tabular}[c]{@{}l@{}}layer2   \\ destination\end{tabular}          & input port & vlan id & ethernet type              & \multicolumn{1}{c|}{\begin{tabular}[c]{@{}l@{}}vlan   \\ priority\end{tabular}} & \begin{tabular}[c]{@{}l@{}}IP   \\ Type-Of-Service\end{tabular} & \multicolumn{1}{c|}{action} \\ \hline
bits        & 32                & 32                & 16                          & 16     & 8        & 48                & 48                & 32      & 12      & 16                    & \multicolumn{1}{c|}{3}     & 6      &                             \\ \cline{1-13}
match & P                 & P                 & R                           & R      & E        & E                 & E                 & E       & E       & E                     & \multicolumn{1}{c|}{E}           & E      &   \\ \hline
R1                        & 181.19.211.54/32  & 98.74.51.88/32    & 67                          & 512    & *        & fa:16:3e:d0:be:01 & fa:16:3e:05:30:c1 & *       & 100     & 0x0800                & *         & *      & \multicolumn{1}{c|}{act0}   \\ \hline
R2                        & *                 & 191.28.225.110    & *                           & 22     & TCP      & *                 & *                 & *       & 56      & 0x0800                & *         & *      & \multicolumn{1}{c|}{act1}   \\ \hline
R3                        & *                 & 191.28.225.110    & *                           & 443    & TCP      & *                 & *                 & *       & 56      & 0x0800                & *                                  & *      & \multicolumn{1}{c|}{act1}   \\ \hline
R4                        & 181.19.21.104/32  & 98.74.67.48/32    & *                           & 512    & *        & fa:16:3e:d5:7e:08 & fa:16:3d:06:60:c5 & 5       & 23      & 0x0800                & *                                  & *      & \multicolumn{1}{c|}{act1}   \\ \hline
R5                        & 180.230.21.104/32 & 13.23.0.0/16      & \multicolumn{1}{l|}{17}     & 21     & UDP      & fa:16:3e:45:6e:11 & fa:16:3e:d0:be:01 & 9       & *       & \multicolumn{1}{l|}{} & *                                  & *      & \multicolumn{1}{c|}{act2}   \\ \hline
R6                        & 185.230.25.204/32 & 45.67.0.0/24      & \multicolumn{1}{l|}{34}     & 512    & *        & fa:16:3e:05:30:c1 & *                 & 11      & 34      & 0x0800                & *                                  & *      & \multicolumn{1}{c|}{act0}   \\ \hline
R7                        & 130.23.0.0/16     & 180.230.21.104/32 & \multicolumn{1}{l|}{0-124}  & 514    & *        & *                 & *                 & *       & 110     & *                     & *                                  & *      & \multicolumn{1}{c|}{act2}   \\ \hline
R8                        & 131.230.0.0/16    & 181.19.21.104/32  & \multicolumn{1}{l|}{0-9800} & 512    & *        & fa:16:3d:06:60:c5 & *                 & *       & 841     & *                     & *                                  & *      & \multicolumn{1}{c|}{act2}   \\ \hline
R9                        & 45.67.8.0/24      & 130.23.0.0/16     & \multicolumn{1}{l|}{34}     & 22     & TCP      & fa:16:3e:05:30:c1 & *                 & 12      & 4095    & *                     & *                                  & *      & \multicolumn{1}{c|}{act0}   \\ \hline
R10                       & *                 & 131.230.0.0/16    & \multicolumn{1}{l|}{*}      & 80     & TCP      & *                 & *                 & 15      & 3200    & 0x8100                & *                                  & *      & \multicolumn{1}{c|}{act3}   \\ \hline
\end{tabular}
\end{table*}

A packet matches a rule if each field in the packet header satisfies the matching condition of the corresponding field in the rule, e.g., the packet's source/destination IP address matches the prefix of the $nw\_src/nw\_dst$ address in the rule, the packet's source/destination port number is contained in the $tp\_src/tp\_dst$ range specified in the rule, and the packet's protocol, source/destination Ethernet addresses, ingress port, Ethernet type, VLAN ID, VLAN priority, IP Type of service matches the rule's $ip\_proto$, $dl\_src \& dl\_dst$, $in\_port$, $eth\_type$, $vlan\_id$, $vlan\_priority$, $IP\_tos$  fields. Rules also can overlap and for this reason, given a packet, it can match a multiple numbers of rules but the classification engine only takes actions defined by the highest priority of rules that packet matches. A sample of 10 such rules is shown in Table~\ref{tab:openflowRuleset}.

There are two major software-based approaches in packet classification literature: decision-tree and decomposition. So far the decision- tree-based approaches rely on heuristics to cut the search space recursively into smaller subspaces~\cite{gupta:99hicuts,singh2003packet}. The fields in the packet header e.g., source and destination IP addresses, source, and destination port numbers, protocol number, source/destination Ethernet addresses, ingress port, Ethernet type, VLAN ID, VLAN priority, IP Type of service all represent the dimensions in the geometric space, a packet is represented as a point in this space, and a rule as a hypercube. Doing packet classification is essentially finding at which hypercube/hypercubes contain that particular point i.e., packet. As the hypercubes could overlap, it is just not only one hypercube that contains the point but also could be multiple of hypercubes contain the same point. From this reasoning, one could say that packet classification is harder or at least equally hard to point-location problem in a d-dimensional geometric space (d is the number of fields considered for packet classification).  In particular, in a $d$-dimensional geometric space with $n$ non-overlapping hypercubes and when $d > 3$, this particular point location problem has either a lower bound of $O(\log{}n)$ time and $O(n^d)$ space or  a lower bound of $O(\log^{d-1}{}n)$ time and $O(n)$ space. From this analysis, it could be said that if we want logarithmic computation time, we need memory that is exponential in the number of dimensions, and if we want linear space, the computation time will be exponential in the logarithm of the number of rules \cite{Gupta:2001}.

A decision-tree-based approach exploits the statistical feature of the ruleset to come up with heuristics that are very much dependent on the ruleset. One heuristic might work for one ruleset but will perform poorly for another different ruleset. As $d$ gets bigger and bigger, even with a great heuristic, decision-tree itself couldn't keep up with increasing wire-speed requirements because of the time and space complexity associated with an increasing number of fields $d$ \cite{Hsieh-manyfield-tnsm-19}.




On the other hand in decomposition-based approaches, the packet headers are split into multiple fields, and lookup operations are performed for all the individual fields independently. The partial results of all the fields are merged to produce the final matching result \cite{Lakshman98high-speedpolicy-based}. If we think carefully what this means is there is a scope to partition the fields into subsets and build individual decision-tree on those subsets. This hybrid approach will allow us to tackle the increasing field requirement into packet classification tasks by simply employing a divide-and-conquer technique. Doing so will enable us to exploit hardware parallelism for traversing each decision-tree concurrently to find out matching candidate rules for each decision-tree and then aggregate the results to come up with the final matching rule. Obviously performing this traversing in parallel, avoids the exponential increase in the time or memory size incurred compared to when performing these operations in a single step.


\subsection{Decision-Tree Algorithms}
Decision-tree-based algorithms consider geometric view for the packet classification problem. Each rule in the ruleset represents a hypercube in  $d$-dimensional space where  $d$ is the number of header fields associated with packet classification. In that convention, each packet is nothing but a point in the  $d$-dimensional space. The decision tree construction algorithm employs different heuristics to cut the space recursively into smaller chunks of spaces. Doing such allows each subspace to end up with fewer rules, which consequently allows a low-cost linear search to find the best matching rule in a leaf-node. After a decision tree is built, the algorithm to classify a packet is straight-forward. Based on the value of the packet header, the algorithm follows the cutting sequence to locate the target subspace (i.e., a leaf node in the decision tree) and then performs a linear search on the rules in this subspace~\cite{Jiang:ToVLSI12}. Here, we briefly summarize two major existing techniques to build decision trees for packet classification: node cutting and rule partition\cite{Liang-NPC-2019}.

\begin{figure*}
 \label{fig:decision-tree}
  \centering
  \includegraphics[width=\linewidth,keepaspectratio,angle=0,scale=1]{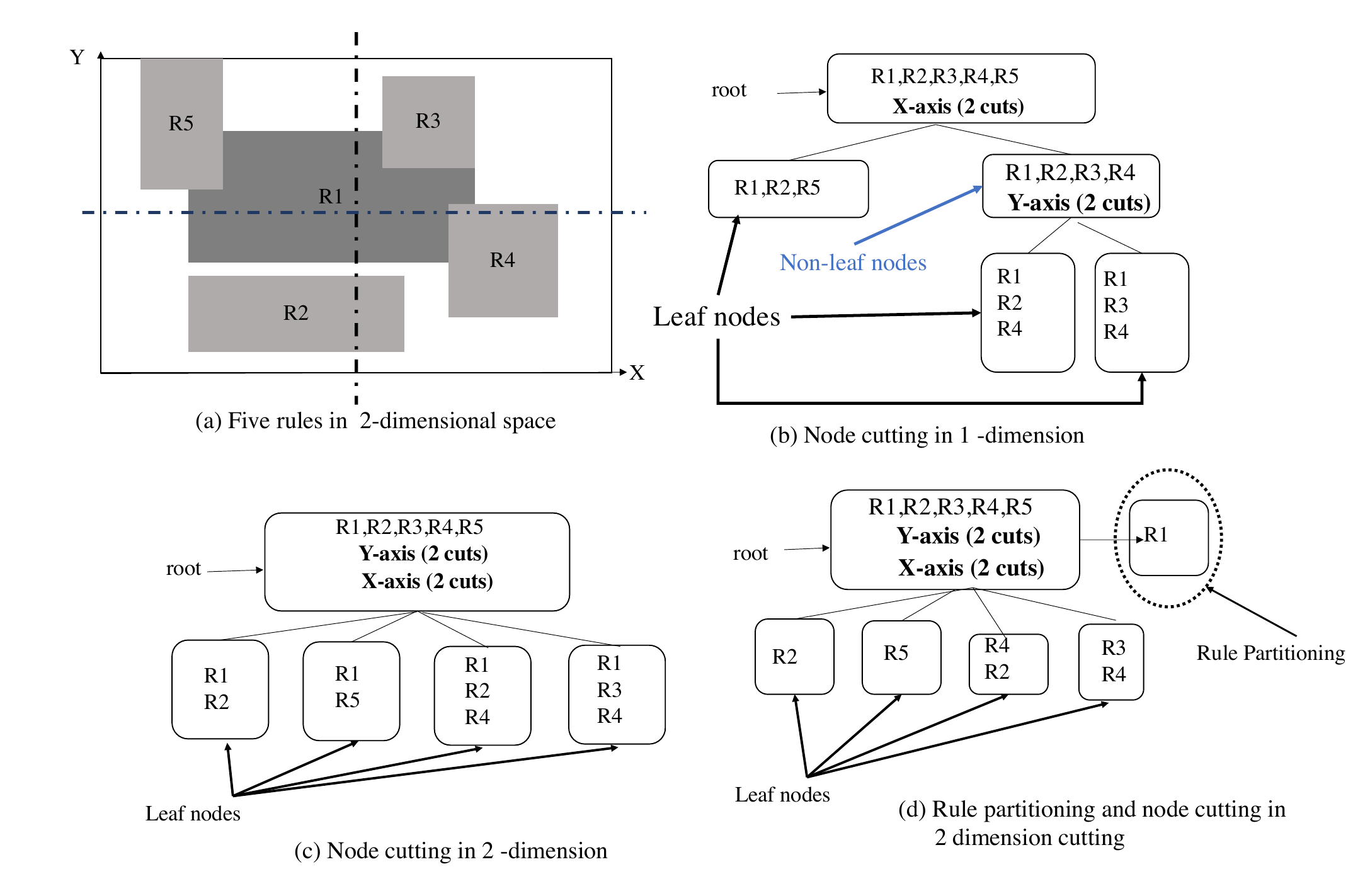}
  \caption{Example of node cutting and rule partitioning process in decision-tree. (a) X and Y axes correspond to any two fields for rules R1–R5 shown as the hypercubes. (b), (c) shows the node cutting process considering one dimension and multi-dimension respectively. (d) shows the common rule in all leaf nodes, i.e., rule R1 is been partitioned so it is not duplicated in all the leaf nodes.}
 
  \label{fig:decision-tree}
\end{figure*}

\subsubsection{Node Cutting}
 Most existing solutions for packet classification aim to build a decision tree that exhibits low classification time (i.e., time complexity) and memory footprint (i.e., space complexity). The main idea is to split nodes in the decision tree by “cutting” along one or more dimensions. Starting from the root which contains all rules, these algorithms iteratively split/cut the nodes until each leaf contains fewer than a predefined number of rules. Given a decision tree, classifying a packet reduces to walk the tree from the root to a leaf, and then chose the highest priority rule associated with that leaf.
Figure \ref{fig:decision-tree} illustrates this technique. The packet classifier contains five rules (R1 to R5) in a two-dimensional space. Figure \ref{fig:decision-tree}(a) shows each rule as a rectangle in the space and represents the cuts as dashed lines. Figure \ref{fig:decision-tree}(b) shows the corresponding decision tree for this packet  classifier. Here first X-axis was cut to create two sub-space (e.g., two nodes) and then Y-axis was cut into two more sub-spaces (e.g., two nodes). The root of the tree contains all the five rules. If a rule intersects a sub-spaces chunk, it is added to that child or that sup-space. For example, R1, R2, and R5 all intersect the first subspace, and thus they are all added to the root’s child node one. If a rule intersects multiple chunks, it is added to each corresponding child, e.g., R1 is added to all the four children. We have another parameter called leaf-threshold, the minimum number of rules in a tree node that will allow the cutting action to stop. For leaf-threshold=3 the left side child of the root node matched the stop criteria. Then, we cut the chunk corresponding right child node along dimension Y. This results in the right child node end up with two more children node. At this point, the stopping criteria get fulfilled for all the nodes and the decision-tree algorithm stops. Figure \ref{fig:decision-tree}(c) shows the node cutting process where instead of cutting 1- dimension at a time, multiple-dimension is considered to cut a node. This results in a lower depth but a wider tree.
\subsubsection{Node Partitioning}
One challenge with ``blindly" cutting a node is that we might end up with a rule being replicated to a large number of nodes. In particular, if a rule has a large size along one dimension, cutting along that dimension will result in that rule being added to many nodes. For example, rule R1 in Figure 2(a) has a large size in dimension $X$ . Thus, when cutting along dimension $X$, R1 will end up being replicated at every node created by the cut. Rule replication can lead to decision trees with larger depths and sizes, which translate to higher classification time and memory footprint.
One solution to address this challenge is to partition rules based on their ``shapes". Broadly speaking, rules with large sizes in a particular dimension are put in the same set. Then, we can build a separate decision tree for each of these partitions. Figure \ref{fig:decision-tree}(d) illustrates this technique. The five rules from the previous are grouped into two partitions. One partition consists of rules R1 and the other partition consists of the remaining rules.

In summary decision trees building process involves two types of actions: node cutting and rule partition. The algorithm answers the following questions and they constitute our reinforcement learning agent's observation and action-space.
\begin{itemize}
\item which node to apply the action?
\item which action to apply (e.g., node cutting or rule partitioning)?
\item how to apply the action (e.g., which dimension(s) to partition or cut and how many cut)?
\item when to stop?
\end{itemize}
\section{Related Work} \label{sec:relatedWork}

Traditional 5-tuple packet classification is a well-studied problem~\cite{Taylor:ACMCS05}. However, the growth of field number in a ruleset poses new challenges regarding system complexity. In this Section, we will first review several 5 fields solutions, and explain that they are not scalable to many field classification due to their complexity. Then we review a few existing heuristics-based many field solutions, which are rulesets dependent. These motivate our work to develop a performance-scalable and ruleset-independent field classification solution. 

\subsection{5-Tuple Algorithmic Solutions}

To increase the system flexibility and decrease the implementation cost, several algorithmic solutions have been discussed for 5-tuple packet classification~\cite{Taylor:ACMCS05}. Tuple space solution~\cite{Srin:Sigcomm99} leverages the fact that the number of distinct bits is much less than the number of rules in a ruleset. The time complexity of tuple space search is the number of hashed memory accesses of tuples in a ruleset. A tuple defines the number of significant bits in a prefix match field, the nesting level and range ID of a range field, and the existence of a value for an exact match field in a ruleset. Tuple-space-based solutions efficiently compress a ruleset by storing those valid bits of each field only. Besides, tuple-space-based solutions perform the search of each tuple independently and take advantage of parallelism. With the growth field number in a ruleset, both tuple number and tuple size increases. A longer processing latency could be expected. Bloom search~\cite{GSwitch_16}  is an improved version of tuple space solution. It uses a bloom filter as an additional stage to filter out the unrelated candidate rules to improve the system performance. According to the discussions in~\cite{Gupta:2001}, the number of tuples for each rule increases with the growth of header fields in a ruleset, up to $O(W^d)$ in the worst case, where $W$ is the size of a field (also called width) and $d$ is the number of header fields (also called dimension). The storage complexity is $O(N)$, where $N$ is the number of rules in a ruleset since each rule is stored exactly one time in the hash table.

Decomposition-based solutions work on each field in a ruleset independently using cross-products~\cite{Taylo05_DCFL}\cite{Srinivasan98_crossproducting} or header chucks~\cite{Gupta:1999}\cite{Lakshman98high-speedpolicy-based} for the intermediate results. These solutions merge the results from different fields to produce the final match results. Since each field is processed individually, more intermediate results will be generated with the growth of field numbers in a ruleset. The increasing merge stages of those additional fields in a ruleset result in a bigger memory requirement and longer processing latency. The system performance is affected by how a solution handles those intermediate results. When the number of fields increases, more intermediate results are generated and are needed to be processed. As discussed in~\cite{Gupta:2001}, the time complexity is $O(dW)$ for cross-product solutions and is $O(d)$ for header chuck solutions. The storage complexity is $O(N^d)$ for~\cite{Taylo05_DCFL}\cite{Srinivasan98_crossproducting}\cite{Gupta:2001} and $O(dN^2)$ for~\cite{Lakshman98high-speedpolicy-based}.\\
Decision-tree-based approaches~\cite{Kennedy_LowP_ToVLSI14}\cite{Lim_BC_2014}\cite{Yang_D2BS_TOC14} analyze all fields in a ruleset to construct tree data structures for an efficient packet header lookup. Tree depth and rule duplication in a decision tree affect the searching efficiency and memory requirement of the implementation. For the matching process, decision-tree-based solutions traverse the tree using field values to make branching decisions at each node until a leaf is reached. Tree depth and rule duplication in a decision tree affect the searching efficiency and memory requirement of one implementation. Both of them increase with the growth of field number which results in an exponential increase of memory requirement and increasing processing latency. According to the discussion in~\cite{Gupta:2001}, the growth of field numbers in the ruleset results in a linear increase of processing latency, and the time complexity is $O(d)$. Based on the nature of decision trees, the rule duplication is carried over to the next layer. The growth of field numbers in the ruleset results in an exponential increase of memory requirement and the storage is $O(N^d)$.


\subsection{Many-field Algorithmic Solutions}
Unlike the well-studied traditional 5-tuple packet classification problem, many-field packet classification is an emerging problem and there are only a few reported solutions. First, a tuple-space-based solution has been used by Open Virtual Switch~\cite{OpenvSwitch_15} to efficiently compress a ruleset by storing the valid bits of each field for its hash functions. Using these hash functions, the proposed solution can perform the search of each tuple independently and take advantage of parallelism. Second, a many-field decomposition-based solution~\cite{Qu_ANCS15} leverages range-tree and hash functions for a 15-field ruleset. The proposed solution divides the input packet header into multiple strides to be processed independently. It then generates the comparison results using a bit vector. Finally, a many-field decision-tree-based solution~\cite{Jiang:ToVLSI12} is proposed for a 12-field ruleset. The proposed solution divides a ruleset into several subsets with individual optimized decision trees as a pipe-line architecture. An improved algorithm~\cite{Qu_TPDS_2016} leverages similar techniques and designs fine-grained processing elements with a 2-dimensional pipelined architecture on FPGA with better performance for a 15-field ruleset.

In summary, most of these solutions are built on heuristics (e.g., increasing split entropy~\cite{gupta:99hicuts}, balancing splits with custom space measures~\cite{gupta:99hicuts}, special handling for wildcard rules~\cite{singh2003packet}) that fails to generalize the process of building a decision tree for a different set of rules. To address the limits of heuristics-based, a learning-based approach is required~\cite{Liang-NPC-2019}. The challenge will be how to design a learning-based method, which will result in a memory and performance scalable packet classification engine~\cite{jamil2020}.

\section{System Overview}\label{sec:sysOverview}

The proposed system architecture is shown in Figure \ref{fig:systemArchitecture}. In the offline stage, in the decomposition phase, the field partition engine takes a ruleset and split out the subset of fields based on the statistical properties present in each field. For each subset of fields, the decision-tree builder builds a separate decision tree. This phase is called a decision-tree building phase. Together this decomposition and decision-tree phases make a hybrid scheme that enables us to build an efficient packet classification engine. In the online stage, an incoming packet header fields are grouped (i.e., field grouping FG) together so subsets generated by the partition engine reflects FG. So at a high level, our propose scheme does the following steps:

\begin{itemize}
\item Partition the ruleset fields into the subset of fields. If $F=\{F_1,...,F_d\}$ is the set of all the fields then partition action creates subset $U$,$V$ such that $F= U \cup V $ and elements of $U$ and $V$, $\{U\},\{V\} \subset F\{F_1,F_2,...,F_d\}$.
\item Building efficient and optimized decision tree for each subset $U$ and $V$ using deep reinforcement learning.
\item In the online phase, for an incoming packet, traverse the band of trees independently of each other and aggregate matched candidate rules from each tree to obtain the final matched rule.
\end{itemize}

The design principles of our systems are as follows:

\begin{itemize}
\item Inspired by traditional fixed 5-tuple packet classification solutions, we adopt a hybrid decomposition-then-decision- tree-based scheme. After decomposing the original many fields into subsets of fields, from each subset, a different decision tree is built. So the number of decision tree build is equal to the number of subsets. We choose decision-tree as the final data structure as decision-tree are regarded as the most scalable packet classification algorithms \cite{Taylor2005}.
\item A deep learning-based decision tree building is proposed. By leveraging the statistical characteristics in a ruleset a deep reinforcement learning agent learns to optimize constructed decision trees for a given ruleset and targeted objectives. This optimization process doesn't rely on hand-tuned heuristics rather rely on the statistical properties present in the given ruleset.
\item We recognize that with 12 fields, a single decision tree suffers from memory explosion. For this reason, we decompose the given 12 fields into multiple subsets so that each one of the subsets contains fewer fields. The resultant trees from those subsets are called a band of trees. We show that each constructed decision tree in the band of trees remains bounded concerning depth and memory requirement.
\item To decompose given fields of a ruleset, we investigate different metrics such as standard deviation (SD) and a novel metric called diversity index (DI). For each field of a given ruleset, we calculate the SD and DI, rank the fields based on their resulting SD and DI and partition them based on the resulting ranks. We also employ random searches in the field space to decompose the fields into subsets. We compare the results obtained from SD and DI based decomposition and random searching based decomposition methods.
\end{itemize}

\begin{figure*}
 \label{fig:systemArchitecture}
  \centering
  \includegraphics[width=\linewidth]{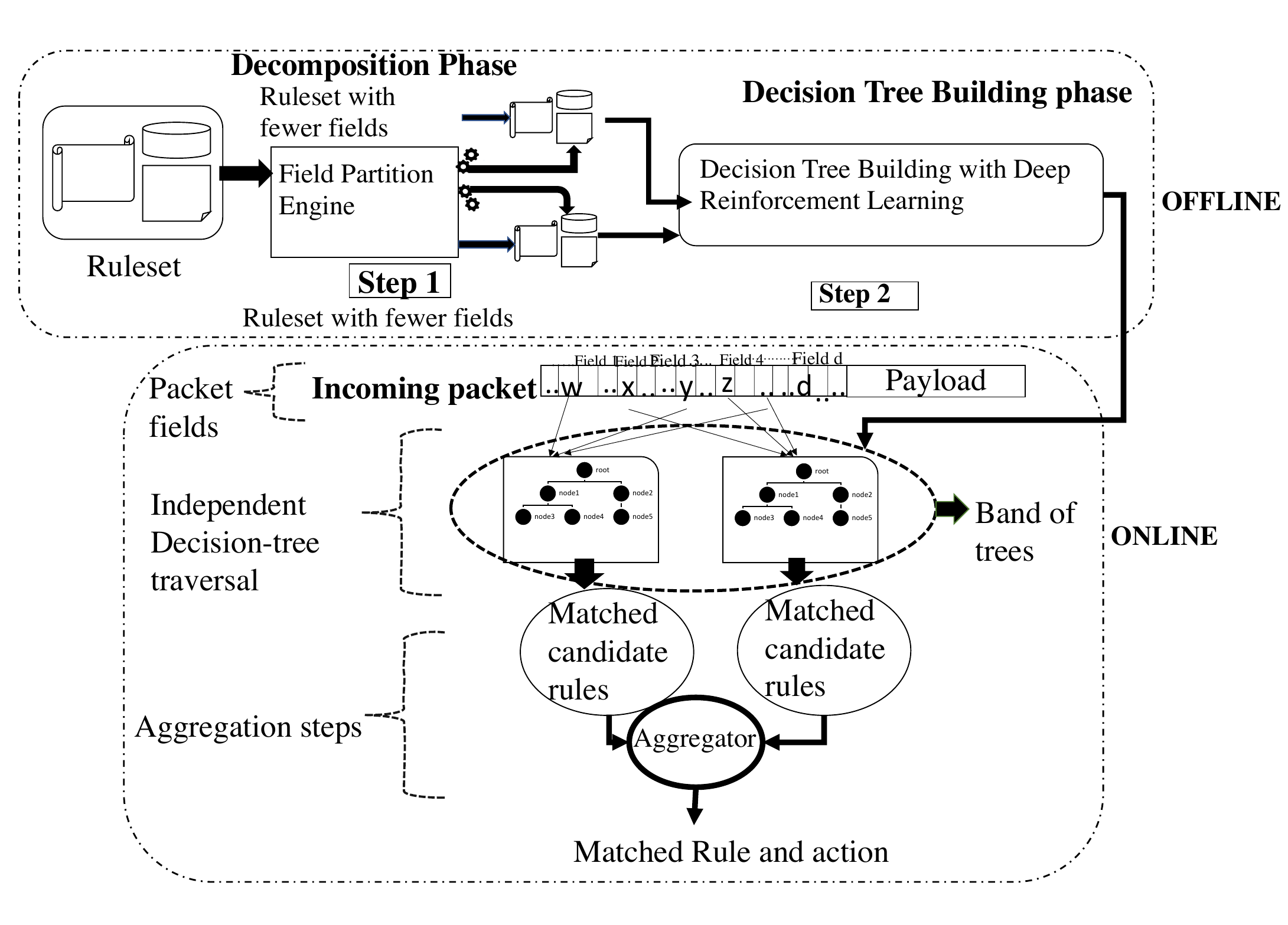}
  
  \caption{System architecture for the proposed packet classification engine. Field Partition engine first partition all the fields into subsets of fields and decision-tree builder builds separate decision trees for each subset of fields. In the online packet classification phase, given an incoming packet,  traversing of decision-trees is done in parallel and independently. Matching candidate rules from trees are combined to get the final matching rule and action.}
  
  \label{fig:systemArchitecture}
\end{figure*}

\textbf{Offline Field Participation using Grouping Metric} To decompose the given fields into different subsets, different statistical measurement of each field on that particular ruleset is considered. The goal of this approach is to generate a more balanced decision tree for each subset. The rationale behind this grouping scheme is that dominant fields (i.e., field that contributes most in building a tree) should be divided into different subsets. For example, if $ip\_src$ and $ip\_dst$ are the most dominant field among all other fields, instead of putting them into one subset, they should be distributed into a different subset and that will enable to build similar depth and size decision-trees for each subset.
The requirement of a similar size decision tree is inspired by the fact that different processor cores could traverse the trees simultaneously and independently and finish at a similar time. If one tree traversal finishes way before that the other one, the system won't be able to take full advantage of hardware parallelism.\\

\textbf{Offline Decision Tree Building Using DRL:} A learning-based system is required to tackle the problem of generating optimal decision trees for a different given environment (i.e., rulesets). As shown in Figure \ref{fig:decisionTreewithRL}, an RL system consists of an agent and an environment, where the agent repeatedly interacts with the environment. In the beginning, environment consists of a set of rules in a root node.  Environment provides the current state $S_t \epsilon S$  which corresponds to the current status of the decision tree. The agent receives this state information and uses a DNN model to choose an action $A_t \epsilon A$, i.e. cut or partition based on a policy. The state and action space are defined in the environment itself. A cut action divides a node along a chosen dimension (i.e., any field of the decomposed field subsets) into some sub-ranges (i.e., 2, 4, 8, 16, or 32 ranges), and creates that many child nodes in the tree.\\
\textbf{Online packet classification:} In the online stage, an incoming packet header fields are grouped (i.e., field grouping FG) together so subsets generated by the partition engine reflects FG. Each field group is then traversed their corresponding decision-trees independently. The candidate rules from each decision-tree traversing are then gone through an aggregator unit to find the matched rule and action.

\section{Building Decision-Tree with Deep Reinforcement Learning}\label{sec:learningMethod}

Most of the decision-tree algorithms rely on heuristics (e.g., increasing split entropy \cite{gupta:99hicuts}, balancing splits with custom space measures \cite{gupta:99hicuts}, special handling for wildcard rules \cite{singh2003packet}) that fails to generalize the process of building a decision tree for a different set of rules. These solutions are specifically tuned to exploit certain characteristics present in a given ruleset and those characteristics may not be present in another ruleset. As a result, this environment-specific heuristics typically suffer from a sub-optimal performance. Another drawback of this hand-tuned heuristics is the absence of a global objective (e.g., tree depth or the number of nodes in the tree). Their decision making is often based on local information (the difference between the number of rules in the current node \cite{Hsieh-manyfield-tnsm-19}, the number of different ranges in different dimensions  \cite{singh2003packet} ). This local information is loosely related to the global objectives and that leads to generating a sub-optimal tree.

\begin{figure}[h]
 \label{fig:decisionTreewithRL}
  \centering
  \includegraphics[width=\linewidth]{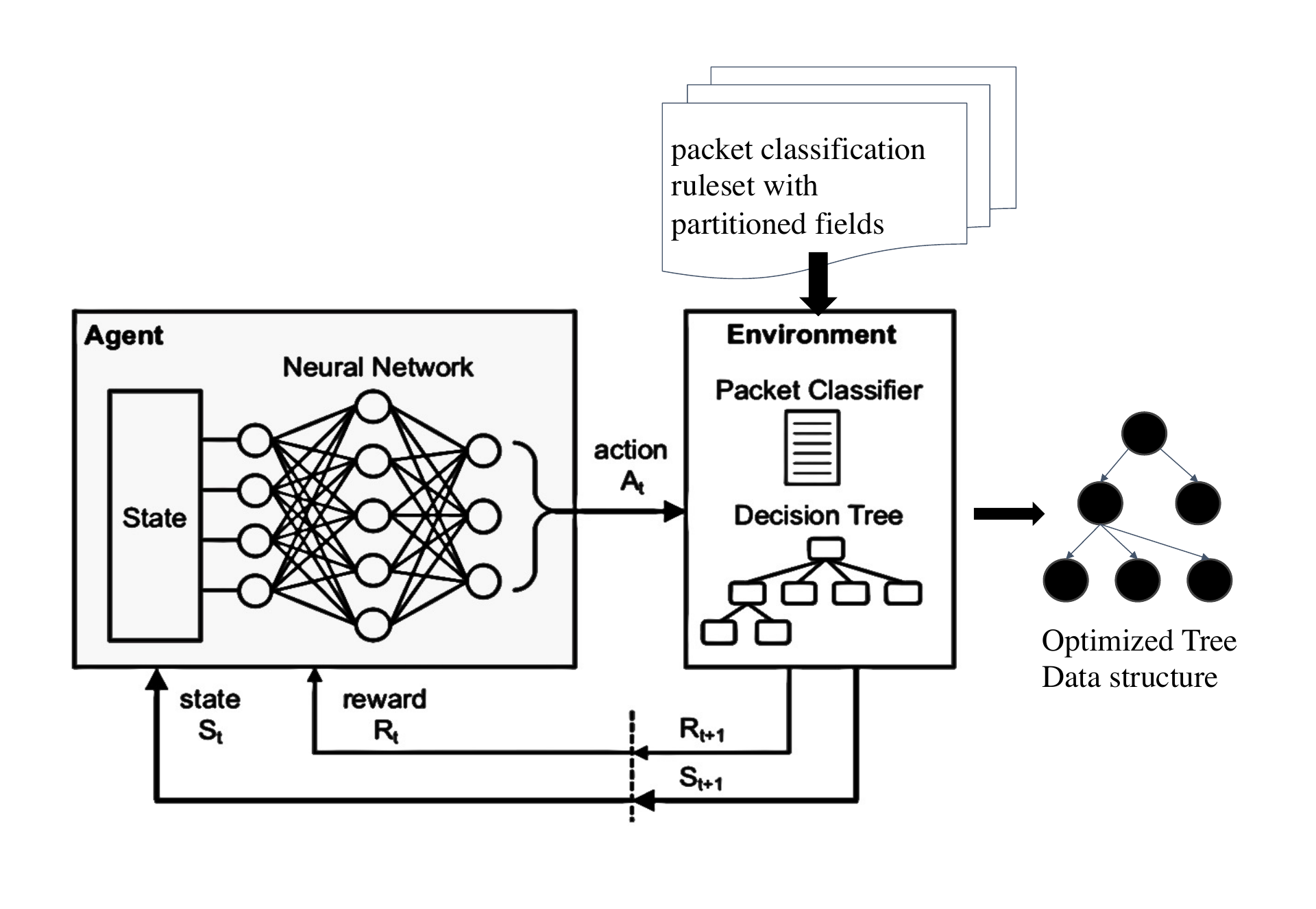}
  \caption{Decision-tree building process with reinforcement learning. The tree itself is the environment, an agent (in this case, a deep neural network) gets the state information from the environment, select an action, and get a reward for that action. If the given action constitutes a good decision-tree as per the objective, the agent receives a big reward and vice versa. The reward-action cycle continues until the optimized tree data-structure has been built.}
  \label{fig:decisionTreewithRL}

\end{figure}

To address above mentioned limits of heuristics-based solutions,  a learning-based approach is needed \cite{Liang-NPC-2019,jamil2020}. The promising aspect of deep learning in systems and networking problems~\cite{MoDong2018,HongziMao2017}, inspires us to use deep learning in the packet classification problem.  In this work, we aim to use a learning-based approach, which has been partly employed in our previous 5-field packet classification~\cite{jamil2020}, however this paper is aiming to address many field packet classification problem, which is even more challenging problem. For benefits of readers, here we briefly summarize the method~\cite{jamil2020} to generate an optimized packet classification engine for any ruleset without relying on heuristics.

A learning-based system is required to tackle the problem of generating optimal decision trees for a different given environment (i.e., rulesets). As shown in Figure \ref{fig:decisionTreewithRL}, an RL system consists of an agent and an environment, where the agent repeatedly interacts with the environment. In the beginning, environment consists of a set of rules in a root node.  Environment provides the current state $S_t \epsilon S$  which corresponds to the current status of the decision tree. The agent receives this state information and uses a DNN model to choose an action $A_t \epsilon A$, i.e. cut or partition based on a policy. The state and action space are defined in the environment itself. A cut action divides a node along a chosen dimension (i.e., any field of the decomposed field subsets) into some sub-ranges (i.e., 2, 4, 8, 16, or 32 ranges), and creates that many child nodes in the tree.

A partition action divides the rules of a node into disjoint subsets (e.g., based on the coverage fraction of a dimension), and creates a new child node for each subset. Depending on the action taken by the agent, the environment also provides a reward signal $R_{t}$. Here, the goal of the model is to learn an optimized single policy  $\pi(a \mid s)$, where $a$ is the action and $s$ is the given state so that the cumulative reward after building the tree is maximized. These steps are repeated at the next time step $(t+1)$ and incrementally build up the tree. As the building process of a tree is deterministic, the tree status in any given time could be encoded and the availability of rewards after a rollout makes the decision tree building process an RL problem. For any action $A$ for any given node $n$ the following expression~\cite{Liang-NPC-2019} is needed to be optimized for a optimize sub-tree rooted at node $n$.

\begin{equation}
V_{n}=\operatorname{argmax}_{a \in A}-\left(c \cdot T_{n}+(1-c) \cdot S_{n}\right),   
\end{equation}
where $T_{n}$ \& $S_{n}$ be the classification time and memory footprint respectively and c is a coefficient. So for every node {n} of the tree, if we optimize $V_{n}$, by induction we optimize the tree itself.

Agent starts with an initial random policy, evaluates this policy with several roll-outs, and then update the policy from the rewards of the roll-outs. A roll-out is a sequence of actions that builds a complete decision tree. All these actions are driven by a policy and a reward is received after completion of building the decision-tree. This process continues until the reward matches the objective value.

One interesting fact that could be leveraged on is that the action on a node entirely depends on the node state itself not the state of the tree. If the subtree rooted at a node could be optimized, recursively the tree rooted from the root node could be optimized (e.g., the memory access time and memory footprint of the tree could be optimized). The worst condition classification time is essentially the height of the tree considering the matching rule is in the farthest leaf node. And the memory footprint is directly related to the number of nodes in the tree.

\begin{figure}[t]
	\centering
	\includegraphics[width=\linewidth]{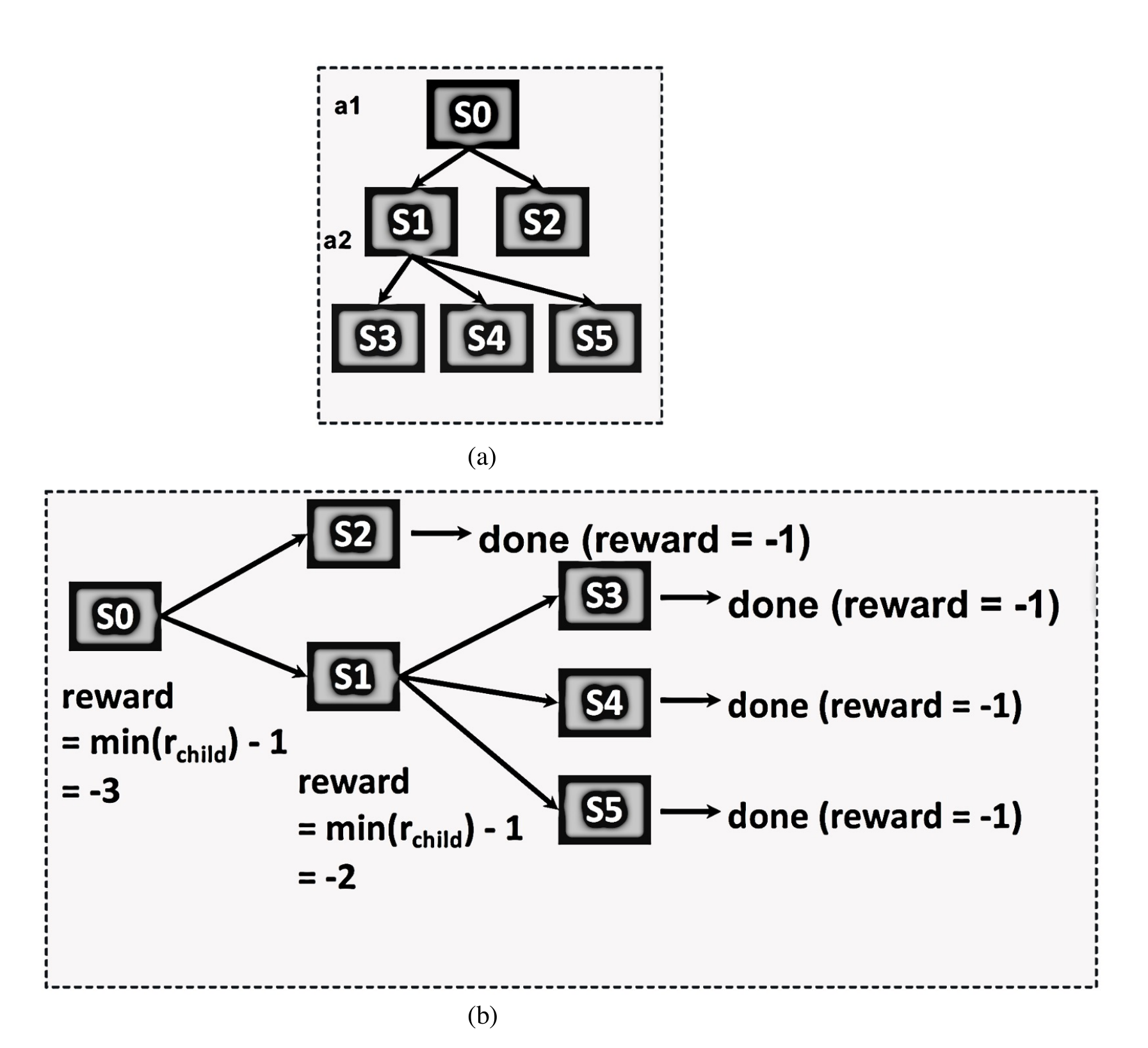}
	\caption{Illustration of reward calculation.(a) shows constructed tree rooted at $S0$ and (b) shows the reward calculation process for the same tree. Reward is calculated for a node once an action is taken and the sub-tree rooted at that node completes. }
	\label{fig:reward-cal}
\end{figure}

The reward signal $(R_t)$ accommodates these two requirements for an action taken to optimize the global objective function of building performance and memory-optimized tree.
In this problem formulation, the environment is considered as a series of 1-step decision problems, each step yielding a reward. We call this secondary award and the actual or primary reward for these 1-step decisions is calculated upon completion of the relevant sub-tree. Calculation of rewards is done not by summing over time but aggregating across tree branches. This is shown in Figure \ref{fig:reward-cal}.  The recursive nature of the decision tree building process allows the reward calculation to be considered as a series of one step decision problem, where each step yields a reward. Once the relevant sub-tree roll out is complete, the actual reward for mentioned one step decisions are calculated. Figure \ref{fig:reward-cal} shows a tree rollout from a root node $s0$. Based on the policy the agent decides to take action $a1$ to split $s0$ into $s1$, $s2$. Of these child nodes $s1$ and $s2$, only $s1$ needs to be further split (via $a2$), into $s3$, $s4$ and $s5$, which eventually finishes the tree. The experiences collected from this rollout consist of two independent 1-step roll outs: ($s0$, $a1$) and ($s1$, $a2$). For a leaf node in the tree if the reward is -1 ,$s2$, $s3$,$s4$ and $s5$ yields -1 each as they are leaf node. Taking the coefficient c = 1 and discount factor = 1 for simplicity, the total reward $R$ for each rollout would be $R$ = -2 and $R$ = -3 respectively for ($s0$, $a1$) and ($s1$, $a2$) following the reward function shown in Figure \ref{fig:reward-cal}. It is important to mention that there is $O(log(n))$  delay between action and reward signal in this approach (where $n$ is the total number of nodes in the tree).

\section{Decomposition of Fields}\label{sec:decomposition}

\subsection{Decomposition based on Standard Deviation ($SD$) and Variance}

Standard deviation measures the relative spread of the values for each field. Standard Deviation for a particular field value is bigger when the differences of that field values are more spread out with respect to distribution mean.
for a set of $N$ numbers $\left\{x_{1}, x_{2}, \cdots, x_{N}\right\}$ if the mean of the collection is $\bar{x}$ then standard deviation, $SD$ is as  follows.\\
\begin{equation}
{SD}=\sqrt{\frac{1}{N-1} \sum_{i=1}^{N}\left(x_{i}-\bar{x}\right)^{2}}.
\end{equation}
Variance was also considered to measure the spread of the fields in a ruleset. $variance$ is described as follows.
\begin{equation}
S^{2}=\frac{\sum\left(x_{i}-\bar{x}\right)^{2}}{N-1},
\end{equation}
$S^{2}=$ sample variance,
$x_{i}=$ the value of the one observation,
$\bar{x}=$ the mean value of all observations,
$N=$ the number of items in the set.

\begin{table}[h]
  \caption[]{$SD$ and variance of $src\_ip$ field for different rulesets. $src\_ip$ field is max Normalized $x^{\prime}=\frac{x}{\max (x)}$ before calculating standard deviation ($SD$) and variance. With decrease in SD and variance of $src\_ip$, the generated tree depth also decreases.}
  \label{tab:sdtable}
  \begin{tabular}{|l|l|c|c|c}
 \cline{1-4}
Ruleset        & SD $src\_ip$ & Variance $src\_ip$ & \multicolumn{1}{c|}{\# of memory access} &  \\ \cline{1-4}
acl3\_1k       & 0.337514   & 0.113916         & 8                                                &  \\ \cline{1-4}
acl3\_1k\_mod  & 0.119811   & 0.014355         & 6                                                &  \\ \cline{1-4}
acl4\_1k       & 0.156345   & 0.024444         & 10                                               &  \\ \cline{1-4}
acl4\_1k\_mod  & 0.119811   & 0.014355         & 6                                                &  \\ \cline{1-4}
acl5\_1k       & 0.270043   & 0.072923         & 9                                                &  \\ \cline{1-4}
acl5\_1k\_mod  & 0.120044   & 0.014411         & 7                                                &  \\ \cline{1-4}
acl5\_10k      & 0.321297   & 0.103232         & 12                                               &  \\ \cline{1-4}
acl5\_10k\_mod & 0.119766   & 0.014344         & 10                                               &  \\ \cline{1-4}
\end{tabular}

\end{table}

To further demonstrate the effect of field variability on tree depth, an experiment with 4 different classbench rulesets has been conducted. Each ruleset has 5 fields ($src\_ip$, $dst\_ip$, $src\_port$, $dst\_port$, and $ip\_proto$) and only the $src\_ip$ field of each ruleset is changed to create another four modified rulesets. So each pair of original and modified ruleset, only the distribution of $src\_ip$ field is different and other four fields $dst\_ip, src\_port, dst\_port, and ip\_proto$ are the same. For example $acl3\_1k$ is original ruleset from classbench \cite{Taylor_ClassBench_2007} and $acl3\_1k\_mod$ is the modified ruleset and the difference between this two ruleset are only in the distribution of first field $src\_ip$. All these rulesets are then used to generate a decision tree with a reinforcement learning agent and the constructed decision tree depth is being observed. Table \ref{tab:sdtable} demonstrate the results. The standard deviation and variance represent the same property, variability of dimension 1 ( $src\_ip$). With a decrease in $SD$ and variance the generated tree depth also decreases as shown in Table~\ref{tab:sdtable}.

\subsection{Decomposition based on Diversity Index ($DI$)}

\begin{table}[t]
\begin{center}
  \caption[]{Sample rulesets to demonstrate construction of Figure~\ref{fig:d-tree-ruleset1-ruleset2}. Only Field 1 of both rulesets has different distribution. Field 2 of both rulesets has same distribution.}.
  \label{tab:diruleset}
  
\begin{tabular}{|c|c|c|c|c|}
\hline

    & field 1    & field 2            & field 1    & field 2            \\ \hline 
$r_1$                 & 200        & 120                & 100        & 120  \\ \hline
$r_2$                 & 180        & 133                & 50        & 133  \\ \hline
$r_3$                 & 185        & 125                & 20        & 125  \\ \hline
$r_4$                 & 189        & 130                & 225        & 130  \\ \hline
$r_5$                 & 187        & 120                & 255        & 120  \\ \hline
$r_6$                 & 186        & 133                & 197        & 133  \\ \hline
$r_7$                 & 170        & 125                & 200        & 125  \\ \hline
$r_8$                 & 172        & 120                & 150        & 120  \\ \hline
$r_9$                 & 174        & 130                & 120        & 130  \\ \hline
$r_{10}$                 & 178        & 120                & 153        & 120  \\ \hline
\end{tabular}

\end{center}
\end{table}

Lets consider two simplified two ruleset  $ruleset 1$ and $ruleset 2$ shown in Table \ref{tab:diruleset}. Lets also assume that from both ruleset, only $field 1$ is considered for building decision trees. $ruleset 1$ , $field 1$ has 10 values and all of them are different. $ruleset 2$ , $field 1$ has 10 values but only 5 of them are distinctive. $Ruleset 1$ is also spread into the space more evenly and range of $ruleset 1$ $field 1$ is higher compared to $ruleset 2$ $field 1$. The normalized range, frequency and density distribution of $field 1$ for $ruleset 1$ , $ruleset 2$ is given in Figure~\ref{fig:frequencyDis}. As $ruleset 2$ has multiple occurrence of same values and range is relatively smaller, $ruleset 2$ is considered more clustered.
Figure \ref{fig:decisionTreewithRL} shows the decision tree building process for $ruleset 1$ and $ruleset 2$. Starting from the root node, every node is cut in half (i.e., midpoint of the attribute range) until the non-leaf node condition is reached. Now considering the decision tree building process for $ruleset 2$, there are several empty child nodes in the tree after the first pass. This is because the $ruleset 2$ $field 1$ values are clustered so a cutting action in a parent node could accommodate a larger number of rules in its child nodes because of the closeness in the values. This is on contrast to $ruleset 1$ where $field 1$ values are more spread.
After removing all the empty child nodes, the resultant pruned decision-tree is achieved for $ruleset 2$, which is shown in the second pass. The tree-height is 7 in the first pass which reduces to 4 after this pruning process. From this experiment, it could be deduced that the frequency of occurrence and the range of values has a direct relationship with the size of the tree. A diversity index is a metric that could capture both of these range and frequency characteristics of a field.

  \begin{figure}[h]
  \label{fig:frequencyDis}
	\centering
	\label{fig:frequency}
	\includegraphics[width=\linewidth]{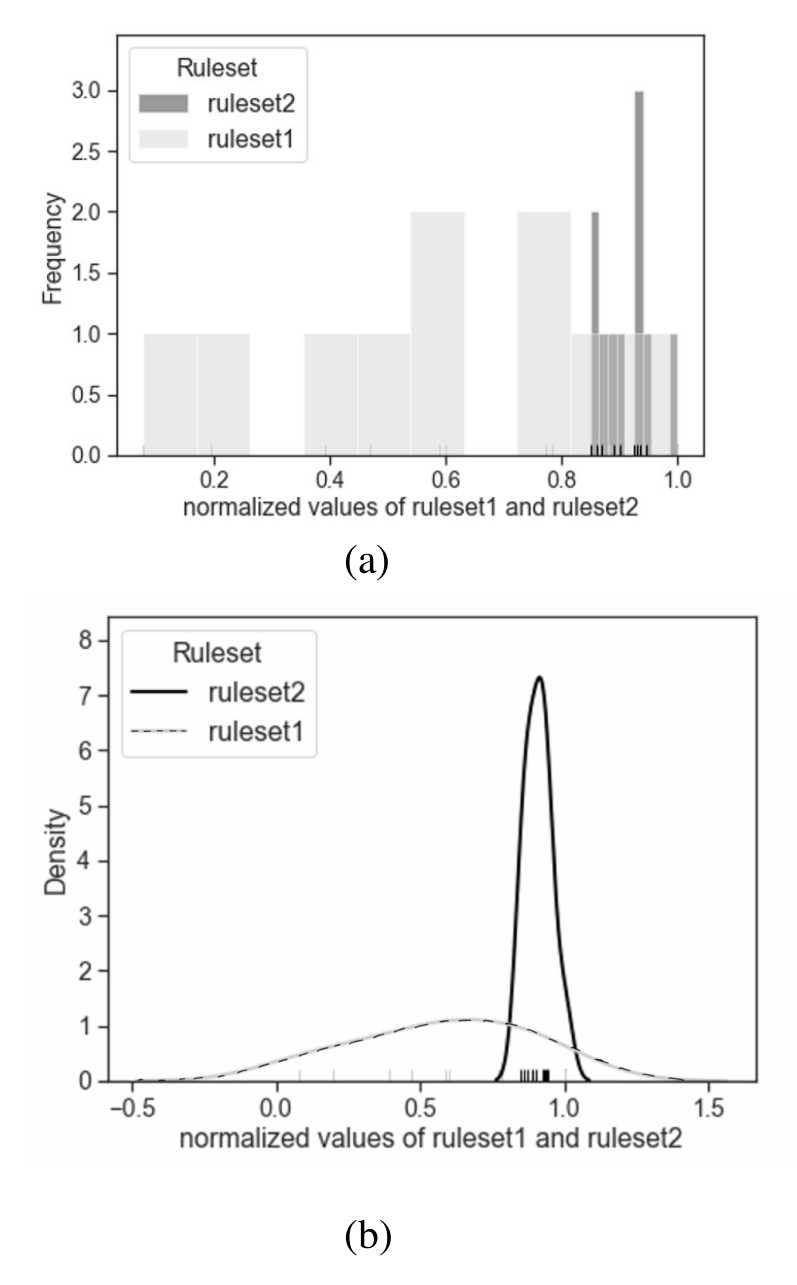}
	\caption{Frequency (a) and density (b) distribution of ruleset1 and ruleset2 with a bin size of 5. Values of $ruleset1$ spreads across the full range (0-1) as values are normalized and frequency of individual values are maximum 1. $ruleset2$ values are congested in between 0.6-1.0 range and frequency of every value is more than 1. The tiny bars are projection of raw data points onto X axis, without involving the binning process of a histogram }
	\label{fig:frequencyDis}

\end{figure}



The derivation of $diversity index $ is  explained in algorithm ~\ref{alg:DiversityIndex}.  \\


\begin{algorithm}
	\SetKwData{Left}{left}\SetKwData{This}{this}\SetKwData{Up}{up}
	\SetKwFunction{Union}{Union}\SetKwFunction{FindCompress}{FindCompress}
	\SetKwInOut{Input}{input}\SetKwInOut{Output}{output}
	\Input{AttributeValues, number of attributes}
	\tcp {AttributeValues is the array of given attribute corresponding values}
	\Output{calculated DiversityIndex of the AttributeValues}
	\BlankLine
	MaxAttributeValue=maximum value present in the input range \\
	\For {every AttributeValues}{
	    normalizedvalue= every attribute Value/MaxAttributeValue;\\
	    add normalized value to normalized attribute value array;\\
	}
	Normalized-No-duplicateValues=remove duplicate values from normalized attribute value array\\
	MaxNormalizedValue=maximum value present in the normalized attribute value array\\
	MinNormalizedValue=minimum value present in the normalized attribute value array\\
	SummationCoefficient=0\\
	\For {every Normalized-No-duplicateValues}{
	    OccuranceValue=number of occurance of every Normalized-No-duplicateValue in normalized attribute value array;
	    NumberOfOccuranceCoefficient=1/OccuranceValue;
	    SummationCoefficient=SummationCoefficient + NumberOfOccuranceCoefficient;
	}
	DiversityIndex=(MaxNormalizedValue-MinNormalizedValue) * SummationCoefficient\\
	\caption{Diversity index calculation algorithm}
	\label{alg:DiversityIndex}
\end{algorithm}


\begin{table}
\begin{center}
\caption[]{DI of $sc\_ip$ field for different rulesets. $sc\_ip$   is max Normalized $x^{\prime}=\frac{x}{\max (x)}$ before calculating diversity index ($DI$). With decrease in $DI$, the generated tree depth also decreases.}
  \label{tab:ditable}
\begin{tabular}{|l|l|l|c}
\cline{1-3}
Ruleset        & DI of source IP & \multicolumn{1}{c|}{\# of memory access} &  \\ \cline{1-3}
acl3\_1k       & 283.943          & 8                                                &  \\ \cline{1-3}
acl3\_1k\_mod  & 0.0398          & 6                                                &  \\ \cline{1-3}
acl4\_1k       & 121.194           & 10                                               &  \\ \cline{1-3}
acl4\_1k\_mod  & 0.03982          & 6                                                &  \\ \cline{1-3}
acl5\_1k       & 169.444          & 9                                                &  \\ \cline{1-3}
acl5\_1k\_mod  & 0.0422          & 7                                                &  \\ \cline{1-3}
acl5\_10k      & 220.204          & 12                                               &  \\ \cline{1-3}
acl5\_10k\_mod & 0.0054           & 10                                               &  \\ \cline{1-3}
\end{tabular}

\end{center}
\end{table}

  \begin{figure*}[]
  \label{fig:d-tree-ruleset1-ruleset2}
	\centering
	\label{fig:d-tree-ruleset1-ruleset2}
	\includegraphics[width=\linewidth]{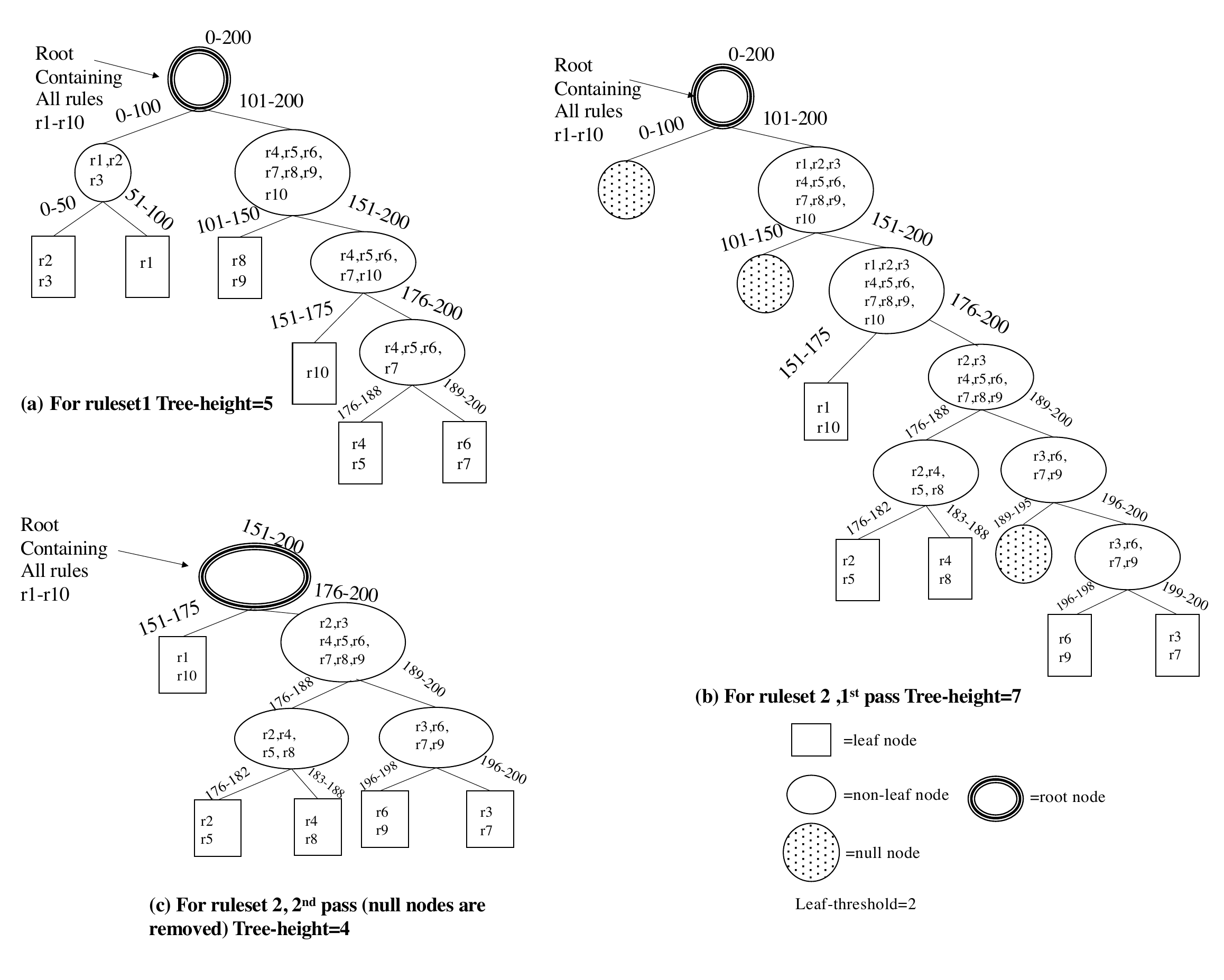}
	\caption{ Decision-tree building process for $ruleset 1$ and $ruleset 2$ in Table \ref{tab:diruleset}. The edges show a range of values associated with each node. (a) is for $ruleset 1$ and (b) \& (c) is for $ruleset 2$. From (a), (b) \& (c), it could be said that if any ruleset field has a lower diversity index it will generate null nodes in the tree building process and those null nodes could be truncated to generate a shorter tree. }
\label{fig:d-tree-ruleset1-ruleset2}

\end{figure*}

First part of line 13 in algorithm ~\ref{alg:DiversityIndex} captures how much range a particular attribute (for example $field1$) has and the second part captures how unique are the values in that attribute. Second part of line 13 in algorithm ~\ref{alg:DiversityIndex} is constructed in such a way that it penalizes by returning a lower value for multiple occurrences of a value. As shown in Figure \ref{fig:d-tree-ruleset1-ruleset2}, the higher the diversity index value for a particular attribute is, the constructed decision tree has higher depth and vice versa. To further demonstrate the effect of $Diversity Index$ on generated tree depth, an experiment with 4 different classbench rulesets has been conducted. Each ruleset has 5 fields ($src\_ip$, $dst\_ip$, $src\_port$, $dst\_port$, and $ip\_proto$) and only the $src\_ip$ field of each ruleset is changed to create another four modified rulesets. For each pair of original and modified ruleset, only the distribution of $src\_ip$ field is different, and the other four fields $dst\_ip, src\_port, dst\_port, and ip\_proto$ are the same. For example $acl3\_1k$ is the original ruleset from classbench and $acl3\_1k\_mod$ is the modified ruleset and the difference between these two is only in the distribution of the first field  $src\_ip$. All these rulesets are then used to generate a decision tree with a reinforcement learning agent and the constructed decision tree depth is being observed. Table \ref{tab:ditable} shows the obtained result. With decrease in $DI$, the generated tree depth also decreases as shown in Table \ref{tab:ditable}.

\begin{figure*}
\label{fig:memoryresults}
\centering
\label{fig:memoryresults}
\includegraphics[width=\linewidth]{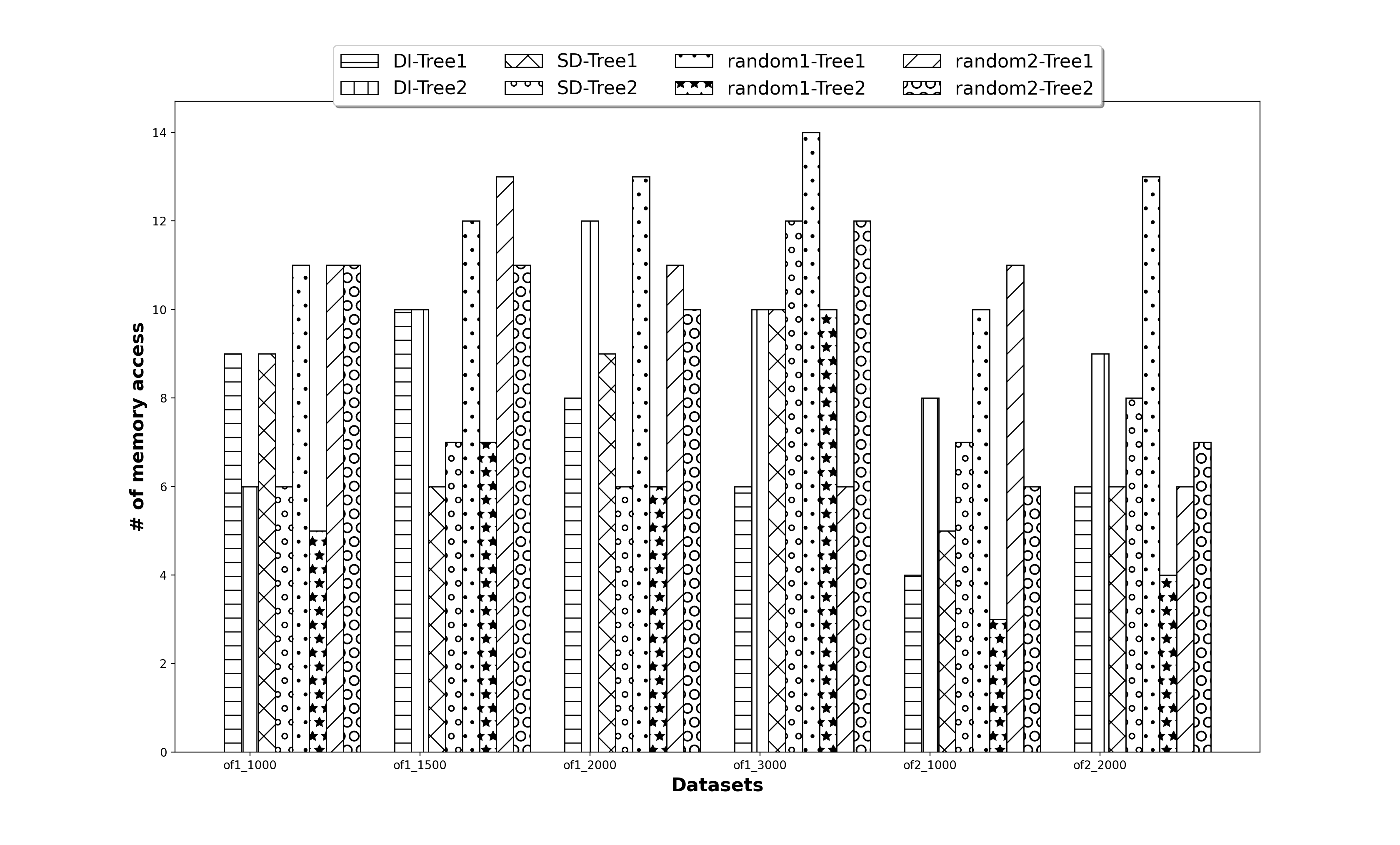}
\caption{ Classification time (tree depth) with DI and SD as partition metric and two random partitioning scheme for 6 classbench-ng generated ruleset. Tree 1  is the tree height obtained from field subset1 and tree2 is from subset2 after decomposing the given fields.}
\label{fig:memoryresults}
\vspace{-2mm}
\end{figure*}

\begin{figure*}
\label{fig:memorysizeresults}
\centering
\label{fig:memorysizeresults}
\includegraphics[width=\linewidth]{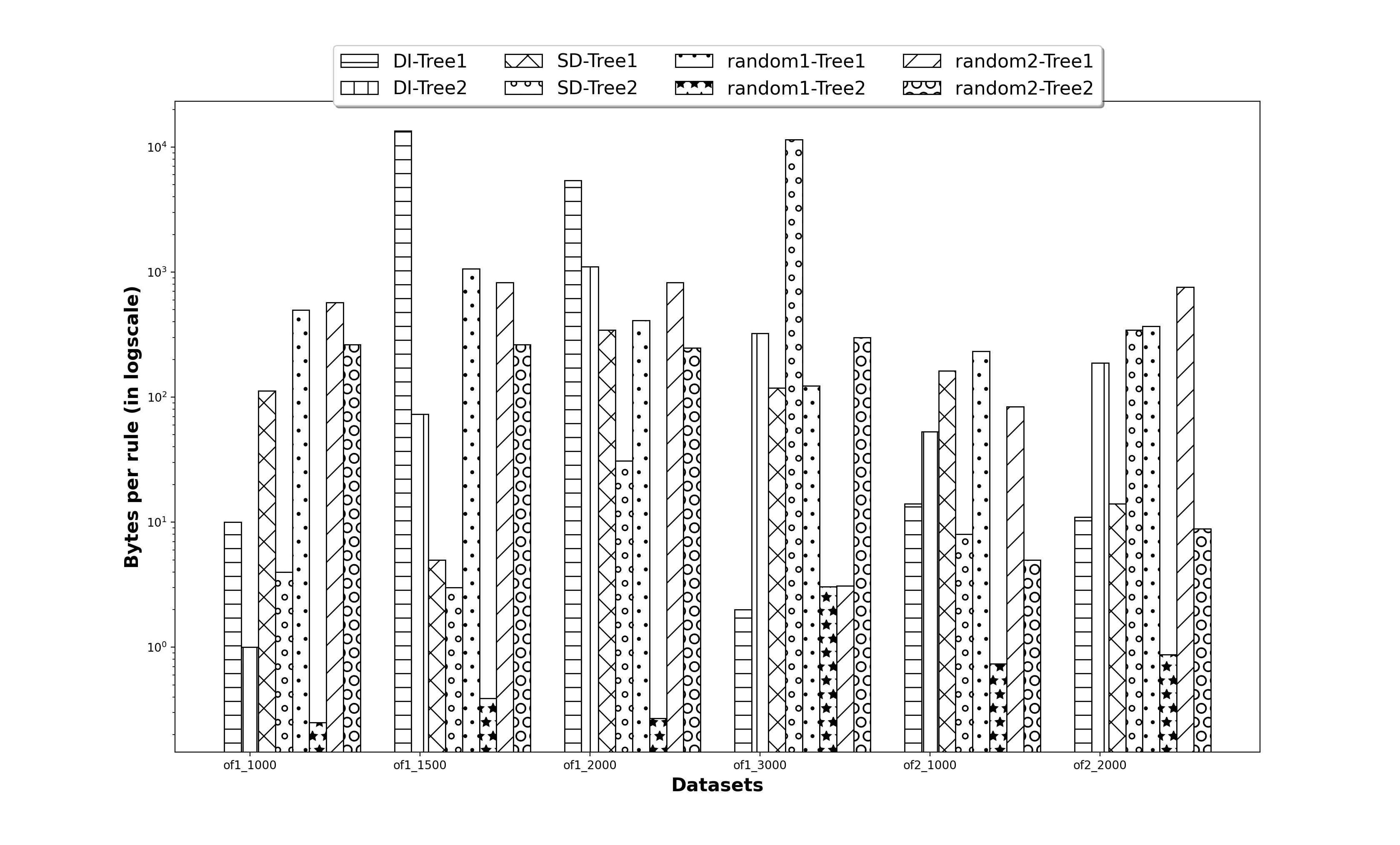}
\caption{ Memory requirement for decision-trees with DI and SD as partition metric and two random partitioning scheme for 6 classbench-ng generated ruleset. Here Tree 1  denotes the memory per byte (in log scale) obtained from field subset1 and tree2 denotes the memory per byte (in log scale) obtained from subset2 after decomposing the given fields. }
\vspace{-2mm}
\label{fig:memorysizeresults}

\end{figure*}

\begin{table*}
\caption[]{$SD$ of different fields for different classbench-ng rulesets. Each field is max Normalized $x^{\prime}=\frac{x}{\max (x)}$ before calculating diversity index ($SD$). The fields are ranked based on their values and odd ranked fields are grouped into one rule subset and even ranked fields into another rule subset.}
  \label{tab:sdtableAllfields}
\tiny
\begin{tabular}{l|ll|ll|ll|ll|ll|ll|}
\cline{2-13}
                                & \multicolumn{2}{c|}{OF1\_1000}       & \multicolumn{2}{c|}{OF1\_1500}       & \multicolumn{2}{c|}{OF1\_2000}       & \multicolumn{2}{c|}{OF1\_3000}     & \multicolumn{2}{c|}{OF2\_1000}       & \multicolumn{2}{c|}{OF2\_2000}       \\ \cline{2-13} 
Field/Ruleset                   & \multicolumn{1}{l|}{Value}    & Rank & \multicolumn{1}{l|}{Value}    & Rank & \multicolumn{1}{l|}{Value}    & Rank & \multicolumn{1}{l|}{Value}  & Rank & \multicolumn{1}{l|}{Value}    & Rank & \multicolumn{1}{l|}{Value}    & Rank \\ \hline
\multicolumn{1}{|l|}{nw\_src}   & \multicolumn{1}{l|}{0.218689} & 3    & \multicolumn{1}{l|}{0.219686} & 3    & \multicolumn{1}{l|}{0.239832} & 2    & \multicolumn{1}{l|}{0.237}  & 3    & \multicolumn{1}{l|}{0.239496} & 2    & \multicolumn{1}{l|}{0.14849}  & 4    \\ \hline
\multicolumn{1}{|l|}{nw\_dst}   & \multicolumn{1}{l|}{0.234153} & 2    & \multicolumn{1}{l|}{0.220896} & 2    & \multicolumn{1}{l|}{0.207661} & 3    & \multicolumn{1}{l|}{0.142}  & 5    & \multicolumn{1}{l|}{0.224667} & 4    & \multicolumn{1}{l|}{0.198316} & 3    \\ \hline
\multicolumn{1}{|l|}{tp\_src}   & \multicolumn{1}{l|}{0.14867}  & 4    & \multicolumn{1}{l|}{0.172602} & 4    & \multicolumn{1}{l|}{0.155446} & 4    & \multicolumn{1}{l|}{0.306}  & 2    & \multicolumn{1}{l|}{0.430966} & 1    & \multicolumn{1}{l|}{0.392906} & 1    \\ \hline
\multicolumn{1}{|l|}{tp\_dst}   & \multicolumn{1}{l|}{0.24017}  & 1    & \multicolumn{1}{l|}{0.252076} & 1    & \multicolumn{1}{l|}{0.255536} & 1    & \multicolumn{1}{l|}{0.334}  & 1    & \multicolumn{1}{l|}{0.226644} & 3    & \multicolumn{1}{l|}{0.235773} & 2    \\ \hline
\multicolumn{1}{|l|}{ip\_proto} & \multicolumn{1}{l|}{0}        & 9    & \multicolumn{1}{l|}{0}        & 9    & \multicolumn{1}{l|}{0}        & 9    & \multicolumn{1}{l|}{0}      & 8    & \multicolumn{1}{l|}{0}        & 9    & \multicolumn{1}{l|}{0}        & 9    \\ \hline
\multicolumn{1}{|l|}{dl\_src}   & \multicolumn{1}{l|}{0.14683}  & 5    & \multicolumn{1}{l|}{0.144232} & 5    & \multicolumn{1}{l|}{0.149172} & 5    & \multicolumn{1}{l|}{0.19}   & 4    & \multicolumn{1}{l|}{0.034381} & 7    & \multicolumn{1}{l|}{0.044307} & 6    \\ \hline
\multicolumn{1}{|l|}{dl\_dst}   & \multicolumn{1}{l|}{0.087401} & 6    & \multicolumn{1}{l|}{0.083561} & 6    & \multicolumn{1}{l|}{0.084258} & 6    & \multicolumn{1}{l|}{0.1}    & 6    & \multicolumn{1}{l|}{0.095486} & 5    & \multicolumn{1}{l|}{0.060709} & 5    \\ \hline
\multicolumn{1}{|l|}{in\_port}  & \multicolumn{1}{l|}{0.004385} & 7    & \multicolumn{1}{l|}{0.043768} & 7    & \multicolumn{1}{l|}{0.037115} & 7    & \multicolumn{1}{l|}{0}      & 9    & \multicolumn{1}{l|}{0.037334} & 6    & \multicolumn{1}{l|}{0.019558} & 7    \\ \hline
\multicolumn{1}{|l|}{vlan\_id}  & \multicolumn{1}{l|}{0}        & 10   & \multicolumn{1}{l|}{0}        & 10   & \multicolumn{1}{l|}{0}        & 10   & \multicolumn{1}{l|}{0}      & 10   & \multicolumn{1}{l|}{0}        & 10   & \multicolumn{1}{l|}{0}        & 10   \\ \hline
\multicolumn{1}{|l|}{eth\_tpe}  & \multicolumn{1}{l|}{0.001645} & 8    & \multicolumn{1}{l|}{0.00163}  & 8    & \multicolumn{1}{l|}{0.001534} & 8    & \multicolumn{1}{l|}{0.0012} & 7    & \multicolumn{1}{l|}{0.001453} & 8    & \multicolumn{1}{l|}{0.0014}   & 8    \\ \hline
\end{tabular}
\end{table*}

\begin{table*}[]
\caption[]{$DI$ of different fields for different classbench-ng rulesets. Each field is max Normalized $x^{\prime}=\frac{x}{\max (x)}$ before calculating diversity index ($DI$). The fields are ranked based on their values and odd ranked fields are grouped into one rule subset and even ranked fields into another rule subset.}
  \label{tab:ditableAllfields }
\tiny
\begin{tabular}{l|ll|ll|ll|ll|ll|ll|}
\cline{2-13}
                                & \multicolumn{2}{c|}{{\color[HTML]{333333} OF1\_1000}}   & \multicolumn{2}{c|}{OF1\_1500}                           & \multicolumn{2}{c|}{OF1\_2000}                          & \multicolumn{2}{c|}{OF1\_3000}                          & \multicolumn{2}{c|}{OF2\_1000}                           & \multicolumn{2}{c|}{OF2\_2000}                          \\ \cline{2-13} 
Field/Ruleset                   & \multicolumn{1}{l|}{Value}                       & Rank & \multicolumn{1}{l|}{Value}                        & Rank & \multicolumn{1}{l|}{Value}                       & Rank & \multicolumn{1}{l|}{Value}                       & Rank & \multicolumn{1}{l|}{Value}                        & Rank & \multicolumn{1}{l|}{Value}                       & Rank \\ \hline
\multicolumn{1}{|l|}{nw\_src}   & \multicolumn{1}{l|}{144.527011}                  & 2    & \multicolumn{1}{l|}{314.51862}                    & 2    & \multicolumn{1}{l|}{510.633542}                  & 2    & \multicolumn{1}{l|}{58.5439533}                  & 3    & \multicolumn{1}{l|}{427.334831}                   & 3    & \multicolumn{1}{l|}{145.078679}                  & 3    \\ \hline
\multicolumn{1}{|l|}{nw\_dst}   & \multicolumn{1}{l|}{81.5808272}                  & 3    & \multicolumn{1}{l|}{63.777913}                    & 4    & \multicolumn{1}{l|}{26.055656}                   & 4    & \multicolumn{1}{l|}{13.399468}                   & 4    & \multicolumn{1}{l|}{43.4939649}                   & 4    & \multicolumn{1}{l|}{68.181425}                   & 4    \\ \hline
\multicolumn{1}{|l|}{tp\_src}   & \multicolumn{1}{l|}{1.05459384}                  & 6    & \multicolumn{1}{l|}{1.01939658}                   & 7    & \multicolumn{1}{l|}{0.17111273}                  & 7    & \multicolumn{1}{l|}{0.25702195}                  & 6    & \multicolumn{1}{l|}{0.07604549}                   & 7    & \multicolumn{1}{l|}{0.03994598}                  & 7    \\ \hline
\multicolumn{1}{|l|}{tp\_dst}   & \multicolumn{1}{l|}{11.7880986}                  & 5    & \multicolumn{1}{l|}{10.335083}                    & 5    & \multicolumn{1}{l|}{9.35056716}                  & 6    & \multicolumn{1}{l|}{5.36726726}                  & 5    & \multicolumn{1}{l|}{580.033114}                   & 2    & \multicolumn{1}{l|}{1484.54923}                  & 2    \\ \hline
\multicolumn{1}{|l|}{ip\_proto} & \multicolumn{1}{l|}{0}                           & 9    & \multicolumn{1}{l|}{0}                            & 9    & \multicolumn{1}{l|}{0}                           & 9    & \multicolumn{1}{l|}{0}                           & 10   & \multicolumn{1}{l|}{0}                            & 9    & \multicolumn{1}{l|}{0}                           & 9    \\ \hline
\multicolumn{1}{|l|}{dl\_src}   & \multicolumn{1}{l|}{40.7218523}                  & 4    & \multicolumn{1}{l|}{71.145901}                    & 3    & \multicolumn{1}{l|}{83.5272569}                  & 3    & \multicolumn{1}{l|}{277.405733}                  & 2    & \multicolumn{1}{l|}{2.56847191}                   & 6    & \multicolumn{1}{l|}{8.23801877}                  & 6    \\ \hline
\multicolumn{1}{|l|}{dl\_dst}   & \multicolumn{1}{l|}{757.741551}                  & 1    & \multicolumn{1}{l|}{1407.00527}                   & 1    & \multicolumn{1}{l|}{1535.53895}                  & 1    & \multicolumn{1}{l|}{2.41*10\textasciicircum{}3}  & 1    & \multicolumn{1}{l|}{1.07*10\textasciicircum{}3}   & 1    & \multicolumn{1}{l|}{2.06*10\textasciicircum{}3}  & 1    \\ \hline
\multicolumn{1}{|l|}{in\_port}  & \multicolumn{1}{l|}{0.12823367}                  & 7    & \multicolumn{1}{l|}{8.98871016}                   & 6    & \multicolumn{1}{l|}{12.0002641}                  & 5    & \multicolumn{1}{l|}{0}                           & 8    & \multicolumn{1}{l|}{4.99546578}                   & 5    & \multicolumn{1}{l|}{4.98642185}                  & 5    \\ \hline
\multicolumn{1}{|l|}{vlan\_id}  & \multicolumn{1}{l|}{0}                           & 10   & \multicolumn{1}{l|}{0}                            & 10   & \multicolumn{1}{l|}{0}                           & 10   & \multicolumn{1}{l|}{0}                           & 9    & \multicolumn{1}{l|}{0}                            & 10   & \multicolumn{1}{l|}{0}                           & 10   \\ \hline
\multicolumn{1}{|l|}{eth\_tpe}  & \multicolumn{1}{l|}{2.62*10\textasciicircum{}-5} & 8    & \multicolumn{1}{l|}{1.44*10\textasciicircum{}-05} & 8    & \multicolumn{1}{l|}{1.48*10\textasciicircum{}-5} & 8    & \multicolumn{1}{l|}{1.33*10\textasciicircum{}-5} & 7    & \multicolumn{1}{l|}{2.59 *10\textasciicircum{}-5} & 8    & \multicolumn{1}{l|}{1.46*10\textasciicircum{}-5} & 8    \\ \hline
\end{tabular}
\end{table*}





In summary, we consider $SD$ and $DI$ as per-ruleset partition metrics to effectively create field subsets. We showed that with a decrease in value of $SD$ and $DI$, the resultant decision-tree size (i.e., depth) also decreases. To generate a similar depth tree from the partitioned ruleset, we can't arbitrarily partition the given ruleset fields and hence couldn't take the complete benefit of hardware parallelization. Apart from $SD$ and $DI$ metric as per ruleset basis partitioning, we also considered a random combination of all the fields across the rulesets in the following result section.

\section{Evaluation}\label{sec:results}
We used python to build the tree environment, that is the tree data-structure and the all-action space and reward calculation steps are constructed into an OpenAI Gym environment. We also used Multi-agent API provided by Ray RLlib library \cite{rllib} which implements parallel simulation and optimization of  RL environments.
Action and observation spaces described in OpenAI Gym format. Actions are sampled from two categorical distributions that select the dimension and action to perform on the dimension respectively. Observations are encoded in a one-hot bit vector that describes the node ranges, partitioning info, and action mask (i.e., for prohibiting partitioning actions at lower levels).We used Proximal policy optimization (PPO)~\cite{SchulmanWDRK17} along with the actor-critic algorithm as described in~\cite{Liang-NPC-2019} to generate optimized trees for the rulesets presented in this section. The hyper-parameters used during the evaluation are given in Table ~\ref{tab:hyperparameters }. 

\begin{table}
\begin{center}
\caption[]{Experimentation hyperparameters. Values in curly braces denote a set of values searched over during evaluation. It is also important to ensure that the rollout timestep limit and model used are sufficiently large for the problem.}
\vspace{1mm}
\label{tab:hyperparameters }
\begin{tabular}{|l|l|}
\hline
Hyperparameter              & Value                                \\ \hline
Time-space coefficient c    & \textless{}user specified\textgreater{} \\ \hline
Max timesteps per rollout   & \{1000, 5000, 15000\}                \\ \hline
Max tree depth              & \{100, 500\}                         \\ \hline
Max timesteps to train      & 10000000                             \\ \hline
Max timesteps per batch     & 60000                                \\ \hline
Model nonlinearity          & fully-connected                      \\ \hline
Model type                  & tanh                                 \\ \hline
Model hidden layers         & {[}512, 512{]}                       \\ \hline
Weight sharing between $\theta,\theta_{v}$ & TRUE                                 \\ \hline
Learning rate               & 0.00005                              \\ \hline
Discount factor            & 1                                    \\ \hline
PPO entropy coefficient     & 0.01                                 \\ \hline
PPO clip param              & 0.3                                  \\ \hline
PPO VF clip param           & 10                                   \\ \hline
PPO KL target               & 0.01                                 \\ \hline
SGD iterations per batch    & 30                                   \\ \hline
SGD minibatch size          & 1000                                 \\ \hline
\end{tabular}

\label{tab:hyperparameters }
\end{center}
\end{table}

ClassBench-ng \cite{classbench-ng} is used to generate packet classifiers with different characteristics and sizes. The size of the rule sets varies from one thousand to three thousand. Our consideration for evaluation is classification time (tree depth) and memory footprint (bytes per rule).
 For different rulesets, all  fields $SD$ and $DI$ values are listed respectively in Tables \ref{tab:sdtableAllfields} and \ref{tab:ditableAllfields }. The values are ranked and based on this rank column, separate subsets are constructed. For example, if we consider ruleset $OF1\_1000$ and \ref{tab:sdtableAllfields}, $nw\_src$, $dl\_src$, $tp\_dst$, $in\_port$, and $ip\_proto$ are grouped into one subset as they occupy odd numbered position in the rank column and remaining fields $nw\_dst$, $dl\_dst$, $tp\_src$, $vlan\_id$, and $eth\_type$ are grouped into different subset as they belong to even numbered position in the rank column. This two subsets are used to generate two separate decision-tree as described in Section 3. $vlan\_priority$ and $ip\_tos$ are not included into the Tables \ref{tab:sdtableAllfields} and \ref{tab:ditableAllfields } because none of the ruleset has any variation for those fields.

 $SD$ and $DI$ are partitioning metrics on a per ruleset basis, that is for each ruleset the field subset members are calculated based on their ranking. As a result, the members are different for the different ruleset. Another cross ruleset partitioning scheme is also following a random walk scheme. In this scheme, for all rulesets, the same fields are grouped and the group members are chosen randomly. Here we only report 2 (i.e., best and worst consecutively) such configurations: \\
 Random 1: $nw\_src$, $nw\_dst$, $tp\_src$, $tp\_dst$, $ip\_proto$ in Subset 1 and
 $dl\_src$, $dl\_dst$, $in\_port$, $vlan\_id$,
 $eth\_type$ in Subset 2.\\
 Random 2 configuration consist of $nw\_src$, $dl\_dst$, $tp\_dst$, $in\_port$, $ip\_proto$ in Subset 1 and
 $dl\_src$, $nw\_dst$, $tp\_src$, $vlan\_id$, $eth\_type$ in Subset 2.\\

 Figure \ref{fig:memoryresults} shows the memory access (i.e., tree depth) for this four configuration. For each configuration and ruleset, we have two trees and their corresponding memory access are stacked on top of each other. Here we try to answer the question of which configuration (i.e., $SD$, $DI$, random1 or random2) results in similar depth trees for subset1 and subset2. From Figure \ref{fig:memoryresults}, we compare performance between four decomposition metrics. Should note the performance of a ruleset is measured by larger depths of two trees built from two decomposed groups. The reported performance of rulesets is the average performance among different rulesets.  The SD decomposition metrics results in 11.5\% faster than DI metrics, 25\% faster than random 2 and 40\% faster than random 1.  This is because SD decomposition produces more balanced trees in terms of tree depth for rule subsets.


 Figure \ref{fig:memorysizeresults} shows the memory required for the generated tree data-structures. For each above-mentioned configuration and ruleset we have two trees and their corresponding memory requirement are stacked on top of each other. It can be seen that not a single configuration achieves a balanced result for all the ruleset but configuration with $SD$ and $DI$ generates overall better memory footprint results for most of the rulesets.

\section{Conclusion}\label{sec:conclusion}

In this work, we present a decomposition and deep reinforcement learning-based solution for many field packet classification problems. Instead of relying on heuristic based algorithms, we present a learning based solution which is a better alternative for many fields presents in OpenFlow-based switches. We leveraged insights obtained from traditional 5-tuple algorithmic solutions and introduce several partition metrics to decompose the ruleset fields into subsets of fields and showed how this subset of fields could be used in building efficient decision-trees. We investigate different decomposition schemes and construct a decision tree for different schemes with deep reinforcement learning and compare the results. The results show that the SD decomposition metrics results in 11.5\% faster than DI metrics, 25\% faster than random 2 and 40\% faster than random 1. One limitation of this work is its evaluation of the small size rulesets.  Nevertheless, we believe that our work is an important step towards learning based many-field scalable packet classification solution.
\bibliographystyle{IEEEtran}
\bibliography{sample-base}


\begin{IEEEbiography}[{\includegraphics[width=0.75in,height=1.1in,clip,keepaspectratio]{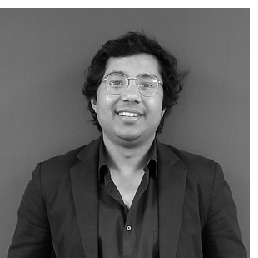}}]{Hasibul Jamil} received his M.S. degree in ECE from Southern Illinois University Carbondale in 2021 and currently a Ph.D. student in Department of Computer Science and Engineering of University at Buffalo,NY. His research interests include machine learning, network systems design, high performance computing, distributed computing systems, performance optimization.

\end{IEEEbiography}
\vskip -2\baselineskip plus -1fil

\begin{IEEEbiography}[{\includegraphics[width=0.75in,height=1.1in,clip,keepaspectratio]{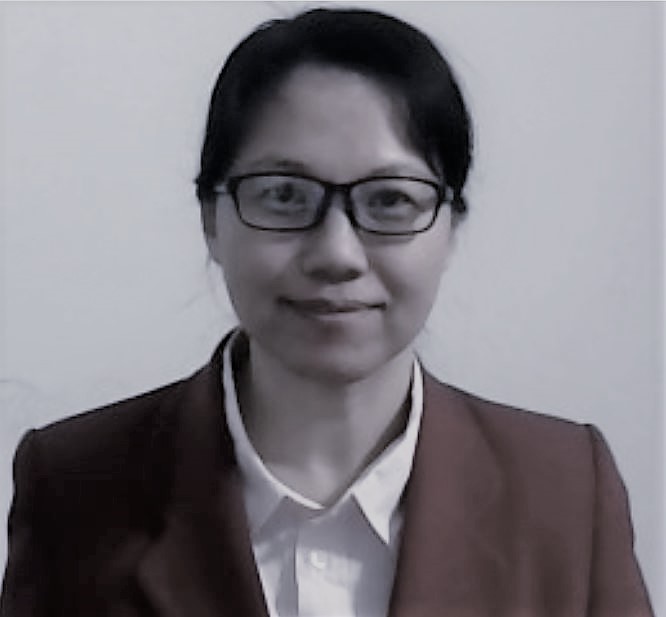}}]{Dr. Ning Yang} is an Assistant Professor in the School of Computing at Southern Illinois University Carbondale, IL. She received a Ph.D. degree in Electrical and Computer Engineering from Southern Illinois University Carbondale in 2020. Her research interests include network security, Internet of Things, future network architectures, and neural networks.
\end{IEEEbiography}
\vskip -2\baselineskip plus -1fil

\begin{IEEEbiography}[{\includegraphics[width=0.75in,height=1in,clip,keepaspectratio]{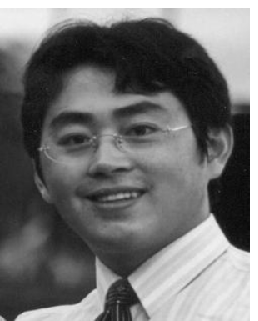}}]{Dr. Ning Weng} is a Full Professor in the School of Electrical, Computer, and Biomedical Engineering at Southern Illinois University Carbondale. He received a Ph.D. degree in electrical and computer engineering from the University of Massachusetts Amherst in 2005. His research interests are in the areas of computer architecture, computer networks, and embedded systems.
\end{IEEEbiography}

\vfill

\end{document}